\documentclass[prb,twocolumn,amsmath,amssymb,superscriptaddress,aps]{revtex4-1}
\usepackage[pdftex]{graphicx}
\usepackage{dcolumn}
\usepackage{amssymb} 
\usepackage{placeins}
\usepackage{tikz}
\usepackage{multirow}
\usetikzlibrary{shapes,arrows}
\usepackage{amsmath}
\usepackage[utf8]{inputenc}
\usepackage{mathtools}
 
\usepackage{bm}
\usepackage{hyperref} 
\graphicspath{{figures/}}
\usepackage{todonotes}

\begin{document}

\title{High throughput computational screening for two-dimensional magnetic materials based on experimental databases of three-dimensional compounds} 

\author{Daniele Torelli}
\affiliation{Computational Atomic-scale Materials Design (CAMD), Department of Physics, Technical University of Denmark, DK-2800 Kgs. Lyngby, Denmark}\author{Hadeel Moustafa}
\affiliation{Computational Atomic-scale Materials Design (CAMD), Department of Physics, Technical University of Denmark, DK-2800 Kgs. Lyngby, Denmark}\author{Karsten W. Jacobsen}
\affiliation{Computational Atomic-scale Materials Design (CAMD), Department of Physics, Technical University of Denmark, DK-2800 Kgs. Lyngby, Denmark}
\author{Thomas Olsen}
\email{tolsen@fysik.dtu.dk}
\affiliation{Computational Atomic-scale Materials Design (CAMD), Department of Physics, Technical University of Denmark, DK-2800 Kgs. Lyngby, Denmark}

\date{\today}

\begin{abstract}
We perform a computational screening for two-dimensional magnetic materials based on experimental bulk compounds present in the Inorganic Crystal Structure Database and Crystallography Open Database. A recently proposed geometric descriptor is used to extract materials that are exfoliable into two-dimensional derivatives and we find 85 ferromagnetic and 61 anti-ferromagnetic materials for which we obtain magnetic exchange and anisotropy parameters using density functional theory. For the easy-axis ferromagnetic insulators we calculate the Curie temperature based on classical Monte Carlo simulations of anisotropic Heisenberg models. We find good agreement with the experimentally reported Curie temperatures of known 2D ferromagnets and identify 10 potentially exfoliable two-dimensional ferromagnets that have not been reported previously. In addition, we find 18 easy-axis anti-ferromagnetic insulators with several compounds exhibiting very strong exchange coupling and magnetic anisotropy.
\end{abstract}

\maketitle

\section{Introduction}
The discovery of two-dimensional (2D) ferromagnetism in 2017\cite{Huang2017a, Gong2017b} has initiated a vast interest in the field of the field. The origin of magnetic order in 2D is fundamentally different from the spontaneously broken continuous symmetry that is responsible for magnetism in three-dimensional materials. In particular, the Mermin-Wagner theorem states that a continuous symmetry cannot be broken at finite temperatures in 2D and magnetic anisotropy therefore becomes a crucial ingredient for magnetic order in 2D. The first report on 2D ferromagnetism involved a monolayer of CrI$_3$,\cite{Huang2017a} which has a strong easy-axis orthogonal to the plane and has a Curie temperature of 45 K. In addition, few-layer structures of CrGeTe$_3$ was reported to exhibit ferromagnetic order down to the bilayer limit.\cite{Gong2017b} However, for the case of a monolayer of CrGeTe$_3$ magnetic order is lost due to the presence of an easy-plane, which comprises a continuous symmetry that cannot be broken spontaneously. Since then several materials have joined the family of 2D magnets. Most notably, CrBr$_3$,\cite{Zhang2019} which have properties very similar to CrI$_3$ but with lower Curie temperatures of 34 K due to smaller magnetic anisotropy, Fe$_3$GeTe$_2$, which is metallic and has a Curie temperature of 130 K\cite{Fei2018}, FePS$_3$\cite{Lee2016} which is anti-ferromagnetic with an ordering temperature of 118 K, and VSe$_2$ where some evidence has been provided for ferromagnetic order at room temperature\cite{Bonilla2018} although the presence of magnetism is being debated\cite{Coelho2019}. In addition, several studies of magnetism in bilayers of various 2D materials have demonstrated that interlayer magnetic coupling can give rise to a plethora of new physical properties.\cite{Sivadas2018, Morell2018, Cardoso2018, Jiang2018, Jiang2018a, Klein2018a, Kim2019a, Kim2019b, Kim2019}

Although the handful of known magnetic 2D materials have been shown to exhibit a wide variety of interesting physics, there is a dire need for discovering new materials with better stability at ambient conditions and higher critical temperatures for magnetic order. Such conditions are not only crucial for technological applications of 2D magnets, but could also serve as a boost for the experimental progress. In addition, the theoretical efforts in the field are largely limited by the few materials that are available for comparison between measurements and calculations. An important step towards discovery of novel 2D materials were taken by Mounet et al.\cite{Mounet2018} where Density Functional Theory (DFT) was applied to search for potentially exfoliable 2D materials in the Inorganic Crystal Structure Database (ICSD) and the Crystallography Open Database (COD). More than 1000 potential 2D materials were identified and 56 of these were predicted to have a magnetically ordered ground state. Another approach towards 2D materials discovery were based on the Computational 2D Materials Database (C2DB),\cite{Haastrup2018, Olsen2018, acsnano.9b06698} which comprises more than 3700 2D materials that have been computationally scrutinized based on lattice decoration of existing prototypes of 2D materials. The C2DB presently contains 152 ferromagnets and 50 anti-ferromagnets that are predicted to be stable by DFT. In addition to these high throughput screening studies there are several reports on particular 2D materials that are predicted to exhibit magnetic order in the ground state by DFT,\cite{Miao2018, Gonzalez2019, Sethulakshmi2019, Kong2019, Kan2013, Akturk2017} as well as a compilation of known van der Waals bonded magnetic materials that might serve as a good starting point for discovering new 2D magnets.\cite{McGuire2017a}

Due to the Mermin-Wagner theorem a magnetically ordered ground state does not necessarily imply magnetic order at finite temperatures and the 2D magnets discovered by high throughput screening studies mentioned above may not represent materials with observable magnetic properties. In three-dimensional bulk compounds the critical temperature for magnetic order is set by the magnetic exchange coupling between magnetic moments in the compound and a rough estimate of critical temperatures can be obtained from mean field theory.\cite{Yosida1996} In 2D materials, however, this is no longer true since magnetic order cannot exist with magnetic anisotropy and mean field theory is always bound to fail. The critical temperature thus has to be evaluated from either classical Monte Carlo simulations or renormalized spin-wave theory of an anisotropic Heisenberg model derived from first principles\cite{Gong2017b, Lado2017, Torelli2019, olsen_mrs} The former approach neglects quantum effects whereas the latter approximates correlation effects at the mean field level. Monte Carlo simulations are not well suited to high-throughput studies, but it has recently been shown that such calculations can be fitted to an analytical expression that is easily evaluated for a given material once the exchange and anisotropy parameters have been computed.\cite{Torelli2019, PhysRevB.100.205409} This approach has been applied to the C2DB resulting in the discovery of 11 new 2D ferromagnetic insulators that are predicted to be stable.\cite{Torelli2019c} In addition 26 (unstable) ferromagnetic materials with Curie temperatures esceeding 400 K have been identified from the C2DB.\cite{Kabiraj2020} However, it is far from obvious that any of these materials can be synthesised in the lab even if DFT predicts them to be stable since they are not derived from experimentally known van der Waals bonded bulk compounds. 

In the present work we have performed a full computational screening for magnetic 2D materials based on experimentally known van der Waals bonded materials present in the ICSD and COD. In contrast to previous high throughput screening of these databases we evaluate exchange and magnetic anisotropy constants for all materials with a magnetic ground state and use these to predict the Curie temperature from an expression fitted to Monte Carlo simulation of the anisotropic Heisenberg model.

\section{Methodology}
The first step in the computational screening is to identify potentially exfoliable 2D structures from the bulk materials present in ICSD and COD. In Ref. \onlinecite{Mounet2018} this was accomplished by identifying layered chemically bonded sub-units followed by a calculation of the exfoliation energy from van der Waals corrected DFT. Here we will instead use a recently proposed purely geometrical method that quantifies the amount of zero-dimensional (0D), one-dimensional (1D), two-dimensional (2D) and three-dimensional (3D) components present in a given material.\cite{Larsen2019} The method thus assigns a 0D, 1D, 2D, and 3D score to all materials and thus quantifies the 0D, 1D, 2D, and 3D character. The scores are defined such that they sum to unity and taking the 2D score $>$ 0.5 thus provides a conservative measure of a material being (mostly) composed of 2D components that are likely to be exfoliable.

The magnetic properties of possible candidate 2D materials are then investigated using first principles Heisenberg models derived from DFT.\cite{Olsen2017, Gong2017b, Lado2017, Torelli2019, olsen_mrs} In particular, if a 2D candidate material has a magnetic ground state we model the magnetic properties by the Hamiltonian
\begin{align}\label{eq:H_ani}
    H=-\frac{J}{2}\sum_{\langle ij\rangle}\mathbf{S}_i\cdot\mathbf{S}_j-\frac{\lambda}{2}\sum_{\langle ij\rangle}S_i^zS_j^z-A\sum_i(S_i^z)^2,
\end{align}
where $J$ is the nearest neighbor exchange coupling, $\lambda$ is the nearest neighbor anisotropic exchange coupling, $A$ is the single-ion anisotropy, and $\langle ij\rangle$ denotes sum over nearest neighbors. $J$ may be positive(negative) signifying a ferromagnetic(anti-ferromagnetic) ground state and we have assumed that the $z$-direction is orthogonal to the atomic plane and that there is in-plane magnetic isotropy. This model obviously does not exhaust the possible magnetic interactions in a material,\cite{Xu2018a} but has previously been shown to provide good estimates of the Curie temperature of CrI$_3$\cite{Lado2017, Torelli2019} and provides a good starting point for computational screening studies.

The thermal properties can then be investigated from either renormalized spin-wave calculations\cite{Tyablikov2013, Yosida1996, Gong2019, Lado2017, Torelli2019} or classical Monte Carlo simulations, \cite{Torelli2019, Lu2019} based on the model \eqref{eq:H_ani} Due to the Mermin-Wagner theorem the magnetic anisotropy constants are crucial for having magnetic order at finite temperatures and for ferromagnetic compounds the amount of anisotropy can be quantified by the spin-wave gap
\begin{equation}\label{eq:delta}
\Delta=A(2S-1)+SN_{nn}\lambda
\end{equation}
where $S$ is the maximum eigenvalue of $S_i^z$ and $N_{nn}$ is the number of nearest neighbors. This expression was calculated by assuming out-of-plane magnetic order and in the present context a negative spin-wave gap signals that the ground state favors in-plane alignment of spins in the model \eqref{eq:H_ani} and implies that the assumption leading to Eq. \eqref{eq:delta} breaks down. Nevertheless, the sign of the spinwave gap comprises an efficient descriptor for the presence of magnetic order at finite temperatures in 2D, since a positive value is equivalent to having a fully broken rotational symmetry in spin-space. 
\begin{figure*}[tb]
	\includegraphics[width = 0.9\textwidth]{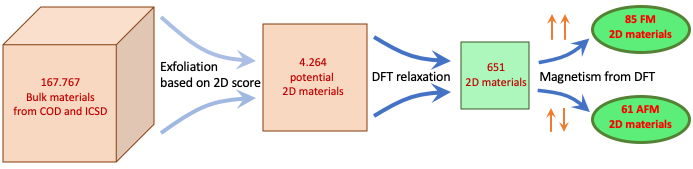}
	\caption{Schematic workflow of the computational discovery of 2D magnets performed in the present work.}
\label{fig:workflow}
\end{figure*}

For bipartite lattices with anti-ferromagnetic ordering ($J<0$) the spinwave analysis based on Eq. \eqref{eq:H_ani} (with out-of-plane easy-axis) yields a spinwave gap of 
\begin{align}\label{eq:delta_afm}
\Delta_\mathrm{AFM}=-\Big[S(J+\lambda)N_{nn}-(2S-1)A\Big]\sqrt{1-\gamma^2},
\end{align}
with 
\begin{align}\label{eq:gamma}
\gamma=\frac{SN_{nn}J}{SN_{nn}(J+\lambda)-(2S-1)A}.
\end{align}
It is straightforward to show that $\Delta_\mathrm{AFM}$ is real and positive if $(2S-1)A>N_{nn}S\lambda$, real and negative if $(2S-1)A<N_{nn}S(2J+\lambda)$ and imaginary otherwise. The latter case corresponds to favouring of in-plane anti-ferromagnetic order and negative real values correspond to favouring of ferromagnetic order (this may happen if $\lambda$ is a large positive number even if $J<0$). $\Delta_\mathrm{AFM}$ thus only represents the physical spinwave gap in the case where it is positive and real. However, in the case of an imaginary spinwave gap the norm of the gap may be used to quantify the strength of confinement to the plane. In the case of non-bipartite lattices we use the expression \eqref{eq:delta_afm} as an approximate measure of the anisotropy. More details on this can be found in Sec. \ref{sec:mapping}.

In Ref. \onlinecite{Torelli2019} it was shown that the critical temperature for ferromagnetic order ($J>0$) can be accurately obtained by classical Monte Carlo simulations of the model \eqref{eq:H_ani}  and for $S>1/2$ the result can be fitted to the function
\begin{equation}\label{eq:tc}
T_\mathrm{C} = \frac{S^2 J T_\mathrm{C}^{\mathrm{Ising}}}{k_\mathrm{B}}f\left( \frac{\Delta}{J(2S-1)} \right )
\end{equation}
where 
\begin{equation}
f(x) = \tanh^{1/4} \left [  \frac{6}{N_{nn}} \log(1+\gamma x) \right]
\end{equation}
and $\gamma = 0.033$. $T_\mathrm{C}^{\mathrm{Ising}}$ is the critical temperature of the corresponding Ising model (in units of $JS^2/k_\mathrm{B}$). The expression \eqref{eq:tc} is readily evaluated for any 2D material with a ferromagnetic ground state once the Heisenberg parameters $J$, $\lambda$ and $A$ have been determined. This can be accomplished with four DFT calculations of ferromagnetic and anti-ferromagnetic spin configurations including spin-orbit coupling. Specifically, for $S\neq 1/2$ the exchange and anisotropy constants are determined by\cite{Torelli2019c, Torelli2020}
\begin{align}
 A &=  \frac{\Delta E_{\mathrm{FM}}(1-\frac{N_{\mathrm{FM}}}{N_{\mathrm{AFM}}})+\Delta E_{\mathrm{AFM}}(1+\frac{N_{\mathrm{FM}}}{N_{\mathrm{AFM}}})}{(2S-1)S},\label{eq:A}\\
 \lambda &=  \frac{\Delta E_{\mathrm{FM}}-\Delta E_{\mathrm{AFM}}}{N_{\mathrm{AFM}}S^2},\label{eq:B}\\
 J &= \frac{ E_{A\mathrm{FM}}^{\parallel}- E_{\mathrm{FM}}^{\parallel}}{N_{\mathrm{AFM}}S^2(1+\beta/2S)},\label{eq:J}
\end{align}
where $\Delta E_{\mathrm{FM}(\mathrm{AFM})}=E_{\mathrm{FM}(\mathrm{AFM})}^{\parallel}-E_{\mathrm{FM}(\mathrm{AFM})}^{\perp}$ are the energy differences between in-plane and out-of-plane magnetization for ferromagnetic(anti-ferromagnetic) spin configurations and $N_{\mathrm{FM}(\mathrm{AFM})}$ is the number of nearest neighbors with aligned(anti-aligned) spins in the anti-ferromagnetic configuration. For bipartite magnetic lattices (square and honeycomb) $N_\mathrm{FM}=0$. However, several of the candidate magnetic materials found below contain a triangular lattice of transition metal atoms and in that case there is no natural anti-ferromagnetic collinear structure to compare with and we have chosen to extract the Heisenberg parameters using a striped anti-ferromagnetic configurations with $N_\mathrm{FM}=2$ and $N_\mathrm{AFM}=4$. Finally the factor of $(1+\beta/2S)$ in the denominator of Eq. \eqref{eq:J} accounts for quantum corrections to anti-ferromagnetic states of the Heisenberg model where $\beta$ is given by 0.202 and 0.158 for $N_\mathrm{AFM}=3$ (honeycomb lattice) and $N_\mathrm{AFM}=4$ (square and triangular lattices) respectively.\cite{Torelli2020} For $S=1/2$ we take $A=0$ and $\lambda=\Delta E_\mathrm{FM}/NS^2$ for $J>0$ and $\lambda=-\Delta E_\mathrm{AFM}/(N_\mathrm{AFM}-N_\mathrm{FM})S^2$ for $J<0$. More details on the energy mapping analysis is provided in appendix A. All DFT calculations were performed with the electronic structure package GPAW\cite{Larsen2017, Enkovaara2010a} including non-selfconsistent spinorbit coupling\cite{Olsen2016a} and the Perdew-Burke-Ernzerhof\cite{pbe} (PBE) functional.

\section{results}
\subsection{Computational screening of COD and ICSD}
The ICSD and COD databases combined count more than 500.000 materials, but removing corrupted or incomplete entries and duplicates, reduces the number to 167767 bulk materials.\cite{Larsen2019} 
Of these, a subset of 4264 are predicted to have a 2D score higher than 0.5 and these materials are the starting point of the present study. We then perform a computational exfoliation by isolating the 2D component and performing a full relaxation of the resulting 2D material with DFT. We restrict ourselves to materials that have a 2D component with less than five different elements and less than a total of 20 atoms in the minimal unit cell. This reduces the number of candidate 2D materials to 651 compounds. We find 85 materials with a ferromagnetic ground state and 61 materials with an anti-ferromagnetic ground state. A schematic illustration of the workflow is shown in Fig. \ref{fig:workflow}.

\begin{figure}[tb]
	\includegraphics[width = 0.5\textwidth]{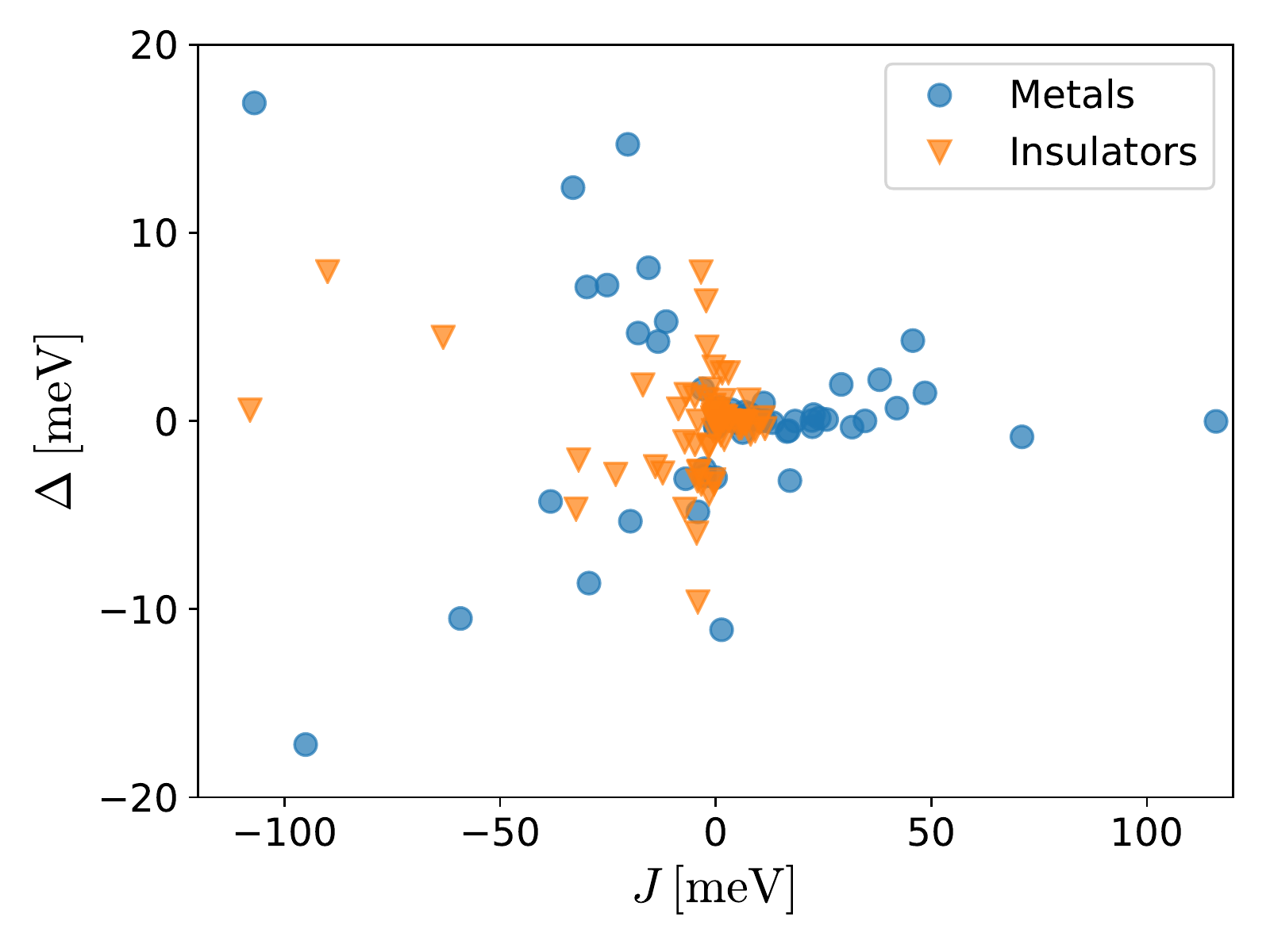}
	\caption{Exchange coupling $J$ and spinwave gap $\Delta$ calculated for the magnetic 2D materials obtained from computational screening of ICSD and COD.}
\label{fig:j_delta}
\end{figure}

For all of the magnetic materials we calculate the exchange coupling $J$ and the spinwave gap $\Delta$ according to the energy mapping approach.\cite{Torelli2019c, Torelli2019e, olsen_mrs} The results are shown in Fig. \ref{fig:j_delta} and all the materials along with the calculated parameters can be found in Tabs. \ref{tab:fm_2el}-\ref{tab:failed}. The spinwave gap is on the order of 0-4 meV for all materials. The exchange couplings fall in the range of 0-10 meV for the insulators but can acquire somewhat larger values for the metals. However, the energy mapping analysis is somewhat ill-defined for metals, since the electronic structure may change significantly when comparing energy differences between ferromagnetic and anti-ferromagnetic configurations. In particular, for insulators the value of $S$ is a well-defined integer that can be extracted from the ferromagnetic ground state without spin-orbit coupling. But for metals it is not clear what value to use in the model \eqref{eq:H_ani}. In addition the Heisenberg model itself is likely to be unsuitable for a description of the magnetic properties of metals and we restrict ourselves to insulators in the following and then subsequently comment on promising metallic compounds.

\subsection{Insulating 2D ferromagnets}
In Tab. \ref{tab:fm_insulators} we display the calculated exchange coupling constants and spinwave gaps for ferromagnetic insulators with $\Delta>0$. Assuming in-plane magnetic isotropy these are the only insulators that will exhibit magnetic order at finite temperatures. For the compounds with $S\neq1/2$ we calculate the Curie temperatures according to Eq. \eqref{eq:tc}. 
\begin{table}[tb]
  \begin{center}
\begin{tabular}{lccrr@{\hskip 0.2in}r}
 {\bf Formula}  & S[$\hbar$] & $N_{nn}$ & $J$ [meV]& $\Delta$ [meV] & $T_\mathrm{C}$ [K] \\
    \hline
NiI$_2$    &   1.0   &   6   &   7.75   &   1.12   &     86 \\ 
CoCa$_2$O$_3$  &   1.5   &   4   &   2.94   &   2.57    &   67 \\ 
CrHO$_2$   &   1.5   &   6   &   2.37   &   0.227   &   45 \\ 
NiRe$_2$O$_8$  &   1.0   &   6   &   1.50   &   2.56    &   31 \\ 
CrI$_3$    &   1.5   &   3   &   1.94   &   1.10    &   28 \\ 
CoCl$_2$    &   1.5   &   6   &   1.85   &   0.0486   &   25 \\ 
VAgP$_2$Se$_6$     &   1.0   &   6   &   2.14   &   0.200    &   21 \\ 
CrBr$_3$   &   1.5   &   3   &   1.84   &   0.276     &   19 \\ 
MnO$_2$  &   1.5   &   6   &   0.508   &   0.434   &      17 \\ 
CrClO    &   1.5   &   6   &   1.04   &   0.0533    &   17 \\ 
CrSiTe$_3$   &   1.5   &   3   &   3.36   &   0.0170   &   15 \\ 
CoCl$_2$O$_8$   &   1.5   &   6   &   0.244   &   0.622   &   10 \\ 
CrCl$_3$    &   1.5   &   3   &   1.29   &   0.0406     &   9.2 \\ 
Mn$_2$FeC$_6$N$_6$  &   2.5   &   3   &   0.102   &   0.155   &   4.4 \\ 
MnNa$_2$F$_3$P$_2$O$_7$  &   1.0   &   2   &   11.0   &   0.182  &   0 \\ 
CuC$_6$H$_4$N$_6$O$_2$     &   0.5   &   2   &   3.04   &   0.0288  &   0 \\ 
MoPO$_5$     &   0.5   &   4   &   0.577   &   0.187   &   -- \\ 
Mn$_3$Cd$_2$O$_8$  &   0.5   &   4   &   0.0625   &   0.470  &   -- \\ 
\end{tabular}
\end{center}
\caption{List of 2D ferromagnetic insulators ($J>0$) with out-of-plane easy axis ($\Delta>0$). The Curie temperature for materials with $S\neq 1/2$ was calculated from Eq. \eqref{eq:tc}.}
\label{tab:fm_insulators}
\end{table}

It is reassuring that the well-known Ising type 2D ferromagnets CrBr$_3$\cite{Zhang2019} and CrI$_3$\cite{Huang2017a} are reproduced by the screening. In addition, CrClO, CrCl$_3$, MnO$_2$, CoCl$_2$, and NiI$_2$ have previously been predicted to be ferromagnetic 2D insulators by DFT.\cite{Haastrup2018, Torelli2019c, Torelli2019e} Multi-layered CrSiTe$_3$ has been reported to exhibit a large magnetic anisotropy in the direction perpendicular to the layers and a ferromagnetic phase transition has been observed at 33 K.\cite{Casto2015} In addition, strained CrSiTe$_3$ has very recently been predicted to comprise an ideal candidate for a 2D Kitaev spin spin-liquid.\cite{Xu2020}

We also find 10 novel 2D ferromagnetic insulators - CoCa$_2$O$_3$, CrHO$_2$, Ni(ReO$_4$)$_2$, Co(ClO$_4$)$_2$, MoPO$_5$, VAgP$_2$Se$_6$, Mn$_2$FeC$_6$N$_6$, MnNa$_2$P$_2$F$_3$O$_7$, Mn$_3$Cd$_2$O$_8$, and CuC$_4$H$_4$N$_2$C$_2$O$_2$N$_4$ that have not been studied prior to the present work. Of particular interest is the compound CoCa$_2$O$_3$, which is predicted to be ferromagnetic up to 57 K. However, it exhibits a rather small band gap of 40 meV, which may imply that the electronic structure could be sensitive to the choice of exchange-correlation functional. Such ambiguities have indeed been reported for FeCl$_3$ and FeBr$_3$, which are both predicted to be small-gap quantum anomalous Hall insulators by PBE, but trivial insulators by PBE+U as well as other GGA functionals.\cite{Olsen2018} 

The largest exchange coupling constant in Tab. \ref{tab:fm_insulators} of 11 meV is found for MnNa$_2$P$_2$F$_3$O$_7$, which appears highly promising. However, we do not have a reliable estimate for the critical temperature due to large in-plane anisotropy (only two nearest neighbors per Mn atom), which renders the inclusion of second nearest neighbors crucial. A faithful estimation of the critical temperature would thus require a full Monte Carlo simulation of an extended Heisenberg model including in-plane anisotropy and exchange couplings. This is, however, beyond the scope of the present screening study.

The materials NiRe$_2$O$_8$ and CoCl$_2$O$_8$ are interesting variants of the common CdI$_2$ prototype (for example NiI$_2$) where the halide atom is replaced by units of ReO$_4$ and ClO$_4$ respectively. For 2D materials discovery based on computational lattice decoration such compounds opens the possibility of a wide range of new materials, since the number of possible ligands in the CdI$_2$ prototype is dramatically increased.

We also wish to mention the compound  CuC$_6$H$_4$N$_6$O$_2$, which is an example of a 2D metal-organic framework (MOF). It is composed of a rectangular lattice of Cu atoms connected by pyrazine (C$_4$H$_4$N$_2$) and C$_2$N$_4$O$_2$ units. Such 2D MOFs have recently attracted an increasing amount of attention and it has been shown that the quasi-2D MOF CrCl$_2$(pyrazine)$_2$ exhibits ferrimagnetic order below 55 K.\cite{Pedersen2018} Due to the spin-1/2 nature of the magnetic lattice we cannot obtain a reliable estimate of the critical temperature of this material. Moreover, the material have large in-plane anisotropy and the second nearest neighbors must play a crucial role since the nearest neighbor approximation gives rise to chains that cannot order themselves at finite temperatures. Nevertheless the sizable value of the intrachain exchange coupling (3.04 meV ) could imply a critical temperature comparable to that of CrI$_3$.

It should be stressed that the results of a screening study like the present one should be taken as a preliminary prediction. The first principles description of magnetic insulators is challenging for DFT since many of these exhibit strong correlation of the Mott-Hubbard type and the calculated Heisenberg parameters may be rather sensitive to the choice of functional.\cite{olsen_mrs, Torelli2019c} A detailed study of the functional dependence or inclusion of Hubbard corrections is required in order to support the theoretical prediction of these 2D materials being ferromagnetic.

\subsection{Itinerant 2D ferromagnets}
For metallic materials the prediction of thermodynamical properties is more challenging since it is not obvious that the Heisenberg Hamiltonian \eqref{eq:H_ani} comprises a good starting point for the analysis. Nevertheless, the exchange coupling $J$ and spin-wave gap $\Delta$ still provides a rough measure of the magnetic interactions and magnetic anisotropy respectively. Alternatively, one could specify the energy difference per magnetic atom in ferromagnetic and anti-ferromagnetic configurations as well as the energy cost of rotating the magnetic moments from the out-of-plane direction to the atomic plane. However, for the sake of comparison we have chosen to report the values of $J$ and $\Delta$ resulting from the energy mapping analysis although it comprises a rather naive approach for metals. The value of $S$ is obtained by rounding off the total magnetic moment per atom to nearest half integer and we then evaluate the critical temperature from Eq. \eqref{eq:tc}, which is the prediction obtained by assuming a Heisenberg model description using the calculated parameters. The results are shown in Tab. \ref{tab:fm_metals}, but it should be kept in mind that the exchange coupling constants and predicted critical temperatures in this case only provides a qualitative measure of the magnetic interactions. 
\begin{table}[tb]
  \begin{center}
\begin{tabular}{lccr@{\hskip 0.2in}rr}
 {\bf Formula} &  S[$\hbar$] & $N_{nn}$ & $J$ [meV]& $\Delta$ [meV]  & $T_\mathrm{C}$ [K] \\
    \hline
FeTe &   1.0   &   4   &   38.0   &   2.19   &     232 \\ 
VCl$_3$    &   1.0   &   3   &   42.0   &   0.679      &   134 \\ 
CrGa$_2$Se$_4$    &   2.0   &   6   &   5.38   &   0.217   &   132 \\ 
CrMoF$_6$   &   1.0   &   4   &   7.84   &   23   &    126 \\ 
NiV$_2$Se$_4$      &   1.0   &   6   &   24.0   &   0.153   &   122 \\ 
FeCl$_2$   &   2.0   &   6   &   4.84   &   0.0454    &   82 \\ 
MnGeMg     &   1.0   &   4   &   11.1   &   0.956    &   75 \\ 
FeBr$_2$  &   2.0   &   6   &   3.24   &   0.0802  &   70 \\ 
VBrO     &   1.0   &   6   &   6.64   &   0.478     &   62 \\ 
CrGa$_2$S$_4$   &   2.0   &   6   &   1.88   &   0.0395    &   39 \\ 
MnSiCa      &   1.0   &   4   &   3.63   &   0.587    &   29 \\ 
FeTaTe$_3$   &   1.0   &   2   &   48.5   &   1.49   &   0 \\ 
CoS$_2$C$_2$N$_2$     &   0.5   &   2   &   25.7   &   0.0765   &   0 \\ 
VFC$_4$H$_4$O$_6$      &   1.0   &   2   &   22.3   &   0.0215   &   0 \\ 
FeCl$_3$    &   0.5   &   3   &   45.7   &   4.27    &   -- \\ 
ScCl   &   0.5   &   9   &   34.6   &   0.00238  &   -- \\ 
FeBr$_3$    &   0.5   &   3   &   29.1   &   1.94    &   -- \\ 
VOBr$_2$  &   0.5   &   4   &   22.7   &   0.336    &   -- \\ 
VS$_2$  &   0.5   &   6   &   11.4   &   0.00854    &   -- \\ 
TiKS$_2$  &   0.5   &   6   &   10.4   &   0.00248    &   -- \\ 
NiLiP$_2$S$_6$   &     0.5   &   6   &   7.90   &   0.0930     &   -- \\ 
FeClO    &   0.5   &   6   &   7.74   &   0.377      &   -- \\ 
Fe$_2$In$_2$Se$_5$    &   0.5   &   9   &   0.867   &   0.711   &   -- \\ 
CoSe  &   0.5   &   4   &   0.247   &   0.0035      &   -- \\ 
\end{tabular}
\end{center}
\caption{List of 2D itinerant ferromagnets ($J>0$ and $E_\mathrm{Gap}=0$) with out-of plane easy axis ($\Delta>0$). The Curie temperature for materials with $S\neq 1/2$ was calculated from Eq. \eqref{eq:tc}.}
\label{tab:fm_metals}
\end{table}

Again, we rediscover a few materials (FeTe and VBrO) that were previously predicted to be ferromagnetic from computational screening of the C2DB. FeClO has recently been exfoliated to bilayer nanoflakes and were shown to retain the anti-ferromagnetic ordering known from the bulk material.\cite{Ferrenti2019} The discrepancy with our prediction of ferromagnetic order could either be due to an inaccurate description by PBE or due to the fact that the true anti-ferromagnetic structure of bulk FeClO is strongly non-collinear,\cite{Grant1971} which is not taken into account in the present simplistic calculations. 

We find a few materials with two nearest neighbors, implying a strongly anisotropic in-plane magnetic lattice. For example, VFC$_4$O$_4$(H$_2$O)$_2$ is a MOF with hydrated alternating linear chains of V and F atoms interconnected by cyclobutanetetrone (C$_4$O$_4$) units. The intra-chain exchange coupling is significant (22.3 meV), but a reliable estimate of the critical temperature requires inclusion of the inter-chain exchange, which is not addressed in the present study. We also find a few materials with 9 nearest neighbors, which originates from a strongly buckled lattice of magnetic atoms and the analysis based on nearest neighbor interactions is expected to be insufficient in this case as well. We observe that several materials have predicted exchange couplings on the order of 10-50 meV, which far exceeds the values found for the insulators. But it should be emphasized that the comparison is not necessarily fair since the electronic structure of the anti-ferromagnetic state may be significantly different compared to the ferromagnetic state. Such differences will lead to large predictions for $J$ that do not originate from magnetic interactions. Nevertheless, Tab. \ref{tab:fm_metals} provides a promising starting point for the discovery of new 2D itinerant ferromagnets, but there is a dire need for a better theoretical framework that can quantitatively deal with the thermodynamical properties of itinerant magnetism in 2D.

We finally note that certain known itinerant 2D ferromagnets (VSe$_2$\cite{Bonilla2018} and CrGeTe$_3$\cite{Gong2017b}) are not present in Tabs. \ref{tab:fm_insulators} and \ref{tab:fm_metals} due to in-plane magnetization, which results in a negative spinwave gap in the present study. For the case of CrGeTe$_3$ this is in accordance with the experimentally observed loss of magnetism in the monolayer limit whereas for VSe$_2$ the origin of magnetic order is still unresolved.\cite{Coelho2019} In addition, we do not find the itinerant 2D ferromagnet Fe$_3$GeTe$_2$,\cite{Fei2018} which cannot be found in a bulk parent form in either the COD or ICSD.

\subsection{Insulating 2D anti-ferromagnets}
In the case of anti-ferromagnetic insulators we do not have a quantitative estimate of the Néel temperature given the nearest neighbor exchange coupling and spin-wave gap. However, it is clear that an easy-axis (positive spinwave gap) is required to escape the Mermin-Wagner theorem for materials with isotropic in-plane magnetic lattices. Moreover, although the formula for the critical temperature Eq. \eqref{eq:tc} was fitted to Monte Carlo simulations we expect that a rather similar expression must be valid for the Néel temperature of anti-ferromagnets. This is partly based on the fact that mean field theory yields similar critical temperatures for ferromagnetic and anti-ferromagnetic interactions in the nearest neighbor model and we thus use the expression \eqref{eq:tc} as a very rough estimate of the critical temperatures for the anti-ferromagnet candidates found in the present work. In Tab. \ref{tab:afm_insulators} we thus display a list of the anti-ferromagnetic insulators with positive spin-wave gap. In addition to the exchange coupling and spin-wave gap we also report the critical temperatures calculated from Eq. \eqref{eq:tc}. 

The most conspicuous result is the exchange coupling of VPS$_3$, which exceeds 0.1 eV. However, while the use of the energy mapping analysis seems to be justified by the gapped anti-ferromagnetic ground state, the ferromagnetic configuration entering the analysis is metallic and may thus imply that the energy difference is not solely due to magnetic interactions. Nevertheless, the local magnetic moments in the ferromagnetic and anti-ferromagnet states are almost identical, which indicates that the large energy difference between the ferromagnetic and anti-ferromagnetic states originates in magnetic interactions.

We also observe that the V and Mn halides are predicted to be anti-ferromagnetic insulators with large exchange coupling constants. However, these compounds exhibits the CdI$_2$ prototype where the magnetic atoms form a triangular lattice. In the present study we have only considered collinear spin configurations, but the true ground state of a triangular lattice with anti-ferromagnetic nearest neighbor exchange has to exhibit a frustrated non-collinear spin structure.\cite{Maksimov2019} Second-nearest neighbors may complicate this picture and the true ground state of these materials could have a complicated structure. Moreover, it has previously been shown that the Mn halides are predicted to be ferromagnetic with the PBE+U functional, which underlines the importance of further investigating the predictions of the present work with respect to exchange-correlation functional, second nearest neighbor interactions etc. 

In analogy with the ferromagnetic insulators NiRe$_2$O$_8$ and CoCl$_2$O$_8$ the anti-ferromagnetic insulator CoRe$_2$O$_8$ comprises a variant of the CdI$_2$ prototype (represented by the V and Mn halides in Tab. \ref{tab:afm_insulators}) where the halide atom has been replaced by ReO$_4$. 

NiC$_2$O$_4$C$_2$H$_8$N$_2$, constitutes an anti-ferromagnetic example of a MOF with a rectangular lattice of Ni atoms connected by a network of oxalate (C$_2$O$_4$) and ethylenediamine (C$_2$H$_4$(NH$_2$)$_2$) units. Again, the material exhibits strong nearest neighbor interactions (across oxalate units), but the second nearest interactions (mediated by ethylenediamine units) will play a crucial role in determining the critical temperature, which is predicted to vanish in the present study, which is only based on nearest neighbor interactions.

Finally, we remark that MnBi$_2$Te$_4$ in 3D bulk form has recently attracted significant attention as it has been demonstrated to comprise the first example of a magnetic Z$_2$ topological insulator.\cite{Otrokov2019, Deng2020} The bulk material is comprised of ferromagnetic layers with anti-ferromagnetic interlayer coupling. In contrast we predict that the individual layers exhibit anti-ferromagnetic order. Like the case of the Mn halides the sign of the exchange coupling constant changes upon inclusion of Hubbard corrections to the DFT description. We have tested that PBE+U calculations yields ferromagnetic ordering for U $>$ 2.0 eV. In addition, we do not find the Ising anti-ferromagnet FePS$_3$,\cite{Lee2016}, since PBE without Hubbard corrections predicts this material to be non-magnetic. This could imply that PBE+U is likely to be a more accurate framework for the present type of calculations, but we leave it to future work to unravel the sensitivity to the choice of xc-functional used for the DFT calculations. 
\begin{table}[tb]
  \begin{center}
\begin{tabular}{lccrrr}
 {\bf Formula}  &  S[$\hbar$] & $N_{nn}$ & $J$ [meV]& $\Delta$ [meV] & $T_\mathrm{C}$ [K] \\
    \hline
VPS$_3$    &   1.0   &   3   &   -108   &   0.58   &   261 \\ 
VBr$_2$    &   1.5   &   6   &   -6.89   &   1.42   &     158 \\ 
ReAg$_2$Cl$_6$    &   1.5   &   6   &   -3.42   &   7.93  &   143 \\ 
VCl$_2$    &   1.5   &   6   &   -4.85   &   1.29   &     119 \\ 
CoRe$_2$O$_8$    &   1.5   &   6   &   -2.22   &   6.39   &  98 \\ 
CoPO$_4$CH$_3$   &   1.5   &   4   &   -2.03   &   3.94   &   56 \\ 
CoSeH$_2$O$_4$     &   1.5   &   4   &   -2.00   &   1.15    &   41 \\ 
MnBr$_2$   &   2.5   &   6   &   -0.576   &   0.322   &    40 \\ 
MnBi$_2$Te$_4$    &   2.5   &   6   &   -0.35   &   0.852  &   35 \\ 
MnCl$_2$   &   2.5   &   6   &   -0.639   &   0.111   &   33 \\ 
MnSH$_2$O$_4$     &   2.5   &   4   &   -0.725   &   0.187    &   28 \\ 
MnSb$_2$F$_12$     &   2.5   &   6   &   -0.292   &   0.251   &   22 \\ 
NiC$_2$O$_4$C$_2$H$_8$N$_2$  &  1.0   &   2   &   -16.9   &   1.92   &    0 \\ 
NbF$_4$    &   0.5   &   4   &   -90.0   &   7.94   &     -- \\ 
VMoO$_5$   &   0.5   &   4   &   -63.2   &   4.45   &  -- \\ 
CuSiO$_3$  &   0.5   &   2   &   -8.66   &   0.644   &    -- \\ 
AgSnF$_6$  &   0.5   &   2   &   -1.16   &   1.71   &   -- \\ 
OsF$_5$KMO   &   0.5   &   2   &   -0.421   &   0.395    &   -- \\ 
\end{tabular}
\end{center}
\caption{List of 2D anti-ferromagnetic insulators ($J<0$) with out-of-plane easy axis ($\Delta>0$). The Curie temperature for materials with $S\neq 1/2$ was calculated from Eq. \eqref{eq:tc}.}
\label{tab:afm_insulators}
\end{table}

\subsection{Itinerant 2D anti-ferromagnets}
For completeness we also display all the predicted anti-ferromagnetic metals in Tab. \eqref{tab:afm_metals}. For $S\neq1/2$, we have provided rough estimates of the critical temperatures based on Eq. \eqref{eq:tc}, but in this case it should be regarded as a simple descriptor combining the effect of exchange and anisotropy rather than an actual prediction for the critical temperature. Neither the energy mapping analysis or the Heisenberg model is expected to comprise good approximations for these materials. However, DFT (with the PBE functional) certainly predicts that these materials exhibit ferromagnetic order at some finite temperature and Tab \eqref{tab:afm_metals} may provide a good starting point for further investigation or prediction of itinerant anti-ferromagnetism in 2D.
\begin{table}[tb]
  \begin{center}
\begin{tabular}{lccrrr}
 {\bf Formula}  &  S[$\hbar$] & $N_{nn}$ & $J$ [meV]& $\Delta$ [meV] & $T_\mathrm{C}$ [K] \\
    \hline
MnAl$_2$S$_4$   &   2.0   &   6   &   -18.0   &   4.67   &   702 \\ 
MnGa$_2$S$_4$   &   2.0   &   6   &   -13.4   &   4.22   &   549 \\ 
MnSnCa   &   2.0   &   4   &   -15.6   &   8.14     &   501 \\ 
MnGeSr   &   1.5   &   4   &   -29.9   &   7.12    &   491 \\ 
MnGeCa    &   1.5   &   4   &   -25.2   &   7.22  &   433 \\ 
MnGeBa   &   2.0   &   4   &   -11.5   &   5.28    &   358 \\ 
MnIn$_2$Se$_4$   &   2.5   &   6   &   -3.03   &   1.70   &   209 \\ 
FeBrSr$_2$O$_3$   &   2.0   &   4   &   -0.153   &   0.614  &   8 \\ 
MnSe$_2$C$_6$N$_4$  &   1.0   &   2   &   -33.1   &   12.4    &   0 \\ 
CrSe  &   0.5   &   4   &   -107   &   16.9    &   -- \\ 
CoI$_2$ &   0.5   &   6   &   -20.4   &   14.7   &   -- \\ 
\end{tabular}
\end{center}
\caption{List of 2D itinerant anti-ferromagnets ($J<0$) with out-of-plane easy axis ($\Delta>0$). The Curie temperature for materials with $S\neq 1/2$ was calculated from Eq. \eqref{eq:tc}.}
\label{tab:afm_metals}
\end{table}

\section{Discussion}
We have performed a computational screening for 2D magnetic materials based on 3D bulk materials present in the ICSD and COD. We find a total of 85 ferromagnetic and 61 anti-ferromagnetic materials, which are listed in Tabs. \ref{tab:fm_2el}-\ref{tab:failed}. The strength of magnetic interactions in the materials have been quantified by the nearest neighbor exchange coupling constants and the magnetic anisotropy has been quantified by the spinwave gap derived from the anisotropic Heisenberg model \eqref{eq:H_ani}. Due to the Mermin-Wagner theorem only materials exhibiting an easy-axis (positive spinwave gap) will give rise to magnetic order at finite temperatures and these materials have been presented in Tabs. \ref{tab:fm_insulators}-\ref{tab:afm_metals}. For these we have also estimated the critical temperature for magnetic order from an expression that were fitted to classical Monte Carlo simulations of the anisotropic Heisenberg model.

The insulating materials are expected to be well described by the Heisenberg model and for $S\neq1/2$ we have evaluated the critical temperatures from an analytical expression fitted to classical Monte Carlo simulations. However, for simplicity this expression was based on a Heisenberg model with in-plane isotropy and nearest neighbor interactions only. This may introduce errors in the prediction of critical temperatures, but for any given material the approach is easily generalized to include other interactions and in-plane anisotropy, which will yield more accurate predictions for critical temperatures. 

A more crucial challenge is related to the determination of Heisenberg parameters from DFT. We have already seen that PBE+U can modify the predictions significantly\cite{Torelli2019c} and even change the sign of the exchange coupling. Is is, however, not obvious that PBE+U will always provide a more accurate prediction compared to PBE (or other exchange-correlation functional for that matter) and benchmarking of such calculations is currently limited by the scarceness of experimental observations.

For anti-ferromagnetic insulators, we expect that classical Monte Carlo simulations combined with the energy mapping analysis will provide an accurate framework for predicting critical temperatures. In the present work we have simply used the expression \eqref{eq:tc} as a crude descriptor and leave the Monte Carlo simulations for anti-ferromagnets to future work. In general, the phase diagrams for anti-ferromagnets will be more complicated compared to ferromagnets\cite{Maksimov2019} and there may be several critical temperatures associated with transitions between different magnetic phases.

The case of itinerant magnets are far more tricky to handle by first principles methods. It is not expected that the applied energy mapping analysis comprises a good approximation for metallic materials and it is not even clear if the Heisenberg description and associated Monte Carlo simulations is the proper framework for such systems. A much better approach would be to use Greens function methods\cite{LIECHTENSTEIN198765, LIECHTENSTEIN1985327} or frozen magnon calculations to access $J(\mathbf{q})\sim\sum_iJ_{0i}e^{i\mathbf{q}\cdot\mathbf{R}_{01}}$ directly from which the magnon dispersion can be evaluated directly. It may then be possible to estimate critical temperatures based on renormalized spinwave theory\cite{Yosida1996} or spin fluctuation theory.\cite{Takahashi2008a}

Despite the inaccuracies in the predicted critical temperatures of the present work, all of the 146 reported magnetic materials constitute interesting candidates for further scrutiny of 2D magnetism. All materials are likely to be exfoliable from bulk structures and contains magnetic correlation in some form. Even the materials with an isotropic magnetic easy-plane that cannot host strict long-range order according to the Mermin-Wagner theorem, may be good candidates for studying KosterLitz-Thouless physics\cite{Thouless1973} Moreover, such materials exhibit algebraic decay of correlations below the Kosterlitz-Thouless transition, which may give rise to finite magnetization for macroscopic flakes.\cite{Holdsworth1994, olsen_mrs} 

\section*{Appendix}
\subsection{Energy mapping analysis}\label{sec:mapping}
Here we provide the details of Eqs. \eqref{eq:A}-\eqref{eq:J} used to extract the Heisenberg parameters from first principles. The energy mapping analysis is based on ferromagnetic and anti-ferromagnetic configurations. We only consider nearest neighbor interactions and in the number of nearest neighbors in the ferromagnetic configurations is denoted by $N$. Only bipartite lattices allow for anti-ferromagnetic configurations where all magnetic atoms have anti-parallel spin alignments with all nearest neighbors. For non-bipartite lattices we thus consider frustrated configurations where each atom has $N_\mathrm{FM}$ nearest neighbors with parallel spin alignment and $N_\mathrm{AFM}$ nearest neighbors with anti-parallel spin alignment. Assuming a {\it classical} Heisenberg description represented by the model \eqref{eq:H_ani}, the ferromagnetic (FM) and anti-ferromagnetic (AFM) DFT energies per magnetic atom with in-plane ($\parallel$) and perpendicular spin configurations are written as
\begin{align}
E_\mathrm{FM}^\perp&=-\frac{(J+B)S^2N_{nn}}{2}-AS^2\\
E_\mathrm{FM}^\parallel&=-\frac{JS^2N_{nn}}{2}\\
E_\mathrm{AFM}^\perp&=\frac{(J+B)S^2(N_\mathrm{AFM}-N_{\mathrm{FM}})}{2}-AS^2\\
E_\mathrm{AFM}^\parallel&=\frac{JS^2(N_\mathrm{AFM}-N_{\mathrm{FM}})}{2},
\end{align}
where $E_0$ represents a reference energy that is independent of the magnetic configuration. The Heisenberg parameters can then be calculated as
\begin{align}
 A &=  \frac{\Delta E_{\mathrm{FM}}(1-\frac{N_{\mathrm{FM}}}{N_{\mathrm{AFM}}})+\Delta E_{\mathrm{AFM}}(1+\frac{N_{\mathrm{FM}}}{N_{\mathrm{AFM}}})}{S^2},\label{eq:At}\\
 \lambda &=  \frac{\Delta E_{\mathrm{FM}}-\Delta E_{\mathrm{AFM}}}{N_{\mathrm{AFM}}S^2},\label{eq:Bt}\\
 J &= \frac{ E_{A\mathrm{FM}}^{\parallel}- E_{\mathrm{FM}}^{\parallel}}{N_{\mathrm{AFM}}S^2},\label{eq:Jt}
\end{align}
where $\Delta E_{\mathrm{FM}(\mathrm{AFM})}=E_{\mathrm{FM}(\mathrm{AFM})}^{\parallel}-E_{\mathrm{FM}(\mathrm{AFM})}^{\perp}$ are the energy differences between in-plane and out-of-plane magnetization for ferromagnetic(anti-ferromagnetic) spin configurations.

However, we wish to base the energy mapping on the quantum mechanical Heisenberg model, which is less trivial. If we start with the anisotropic Heisenberg model where spin-orbit coupling is neglected the ferromagnetic configuration with energy $E_\mathrm{FM}$ corresponds to an eigenstate with energy $-J/2N_\mathrm{AFM}$ per magnetic atom, which is the same as the classical Heisenberg model. However the anti-ferromagnetic configuration does not correspond to a simple eigenstate of the Heisenberg model. In particular, for bipartite lattices the Neel state where all sites host spin that are eigenstates of $S_z$ is not the eigenstate of lowest(highest) energy of the Heisenberg Hamiltonian model with $J<0(J>0)$. Rather the classical energy corresponds to the expectation value of the Heisenberg Hamiltonian with respect to this state whereas the true ground state has lower(higher energy) leading to an overestimation of $J$ if the energy mapping is based on the classical Heisenberg model. We have recently shown how to include quantum corrections to $J$ for bipartite lattices using a correlated state, which has an energy in close proximity to the true anti-ferromagnetic ground state.\cite{Torelli2020} We note that the magnetic moments obtained with DFT support the fact that the DFT energy of the  anti-ferromagnetic configuration represents a proper eigenstate of the Heisenberg model rather than the classical state. The result is the factor of $(1+\beta/2S)$ in equation \eqref{eq:J}. 

Including spin-orbit coupling and magnetic anisotropy in the energy mapping complicates the picture since only one of the states $E_\mathrm{FM}^\parallel$, $E_\mathrm{FM}^\perp$ represents an eigenstate of the anisotropic Heisenberg model. On the DFT side this is reflected by the fact that only one of these configurations would be obtainable as a self-consistent solution and we have to calculate these energies by including spin-orbit coupling non-self-consistently. We thus retain the classical expression for the anisotropy constants, but retain the quantum correction for the exchange constants. Is is, however, clear that the single-ion anisotropy term becomes a constant for any system with $S=1/2$. Since $A$ does not have any physical significance it cannot influence the values of $E_\mathrm{FM(AFM)}^\parallel$ and $E_\mathrm{FM(AFM)}^\perp$ and we take $A=0$ and $\lambda=\Delta E_\mathrm{FM}/NS^2$ for $J>0$ and $\lambda=-\Delta E_\mathrm{AFM}/(N_\mathrm{AFM}-N_\mathrm{FM})S^2$ for $J<0$. In principle, the two choices for $\lambda$ should be equivalent and we have tested that they yield nearly the same value for a few spin-1/2 insulators. But in order to obtain full consistency with the spinwave gap we use different expressions depending on the sign of $J$. In addition for $S\neq1/2$ the classical analysis leads to an inconsistency since the spinwave gap \eqref{eq:delta} is not guaranteed to yield the same sign as $-\Delta E_\mathrm{FM}$. This can be fixed by taking $2S\rightarrow(2S-1)S$ in Eq. \eqref{eq:At}, which leads to Eq. \eqref{eq:A}. Finally, the anti-ferromagnetic spinwave gap Eq. \eqref{eq:delta_afm} was derived for bipartite lattices and it is not possible to derive a gap for non-bipartite lattices in a collinear spin configuration, since such a state will not represent the ground state leading to an instability in the gap. However, we will apply the expression naively to non-bipartite lattices as well but taking $N_{nn}\rightarrow N_\mathrm{AFM}-N_\mathrm{FM}$ to ensure that the sign of the gap corresponds to the sign of $-\Delta E_\mathrm{AFM}$.

\subsection{List of predicted magnetic materials}\label{sec:tables}

\begin{table*}[tb]
  \begin{center}
\begin{tabular}{lr@{\hskip 0.2in}r@{\hskip 0.2in}r@{\hskip 0.2in}rrr@{\hskip 0.2in}rr}
  {\bf Stoichiometry} & {\bf ID} & $S$ [$\hbar$] & $N_{nn}$ & $J$ [meV] & $\Delta$ [meV] & $E_\mathrm{Gap}$ [eV] & $T_\mathrm{C}$ \\
  \hline
ScCl   &   4343683   &   0.5   &   9   &   34.6   &   0.00238   &   0.00   &   -- \\ 
VSe   &   162898   &   0.5   &   4   &   1.61   &   -0.0151   &   0.00   &   0 \\ 
FeTe   &   44753   &   1.0   &   4   &   38.0   &   2.19   &   0.00   &   232 \\ 
CoSe   &   162902   &   0.5   &   4   &   0.247   &   0.0035   &   0.00   &   -- \\ 
YCl   &   4344519   &   0.5   &   9   &   5.29   &   -0.100   &   0.00   &   0 \\ 
YI   &   151974   &   0.5   &   9   &   31.6   &   -0.330   &   0.00   &   0 \\ 
ScO$_2$   &   9009156   &   0.5   &   6   &   5.97   &   -0.017   &   0.73   &   0 \\ 
TiCl$_3$   &   29035   &   0.5   &   3   &   116   &   -0.0233   &   0.00   &   0 \\ 
VS$_2$   &   86519   &   0.5   &   6   &   11.4   &   0.00854   &   0.00   &   -- \\ 
VSe$_2$   &   1538289   &   0.5   &   6   &   22.4   &   -0.311   &   0.00   &   0 \\ 
VI$_2$   &   246907   &   1.5   &   6   &   0.332   &   -0.017   &   1.21   &   0 \\ 
MnO$_2$   &   9009111   &   1.5   &   6   &   0.508   &   0.434   &   1.13   &   17 \\ 
FeCl$_2$   &   9009128   &   2.0   &   6   &   4.84   &   0.0454   &   0.00   &   82 \\ 
FeBr$_2$   &   8101148   &   2.0   &   6   &   3.24   &   0.0802   &   0.00   &   70 \\ 
CoO$_2$   &   20566   &   0.5   &   6   &   16.5   &   -0.563   &   0.00   &   0 \\ 
CoCl$_2$   &   9008030   &   1.5   &   6   &   1.85   &   0.0486   &   0.36   &   25 \\ 
CoBr$_2$   &   9009099   &   1.5   &   6   &   1.20   &   -0.715   &   0.34   &   0 \\ 
NiCl$_2$   &   2310380   &   1.0   &   6   &   6.60   &   -0.00573   &   1.22   &   0 \\ 
NiBr$_2$   &   9009131   &   1.0   &   6   &   6.77   &   -0.0888   &   0.87   &   0 \\ 
NiI$_2$   &   9011538   &   1.0   &   6   &   7.75   &   1.12   &   0.43   &   86 \\ 
CdO$_2$   &   23415   &   1.0   &   6   &   71.0   &   -0.846   &   0.00   &   0 \\ 
VCl$_3$   &   1536707   &   1.0   &   3   &   42.0   &   0.679   &   0.00   &   134 \\ 
CrCl$_3$   &   1010575   &   1.5   &   3   &   1.29   &   0.0406   &   1.75   &   9.2 \\ 
CrBr$_3$   &   1010151   &   1.5   &   3   &   1.84   &   0.276   &   1.52   &   19 \\ 
CrI$_3$   &   251655   &   1.5   &   3   &   1.94   &   1.10   &   1.27   &   28 \\ 
FeCl$_3$   &   1535681   &   0.5   &   3   &   45.7   &   4.27   &   0.00   &   -- \\ 
FeBr$_3$   &   76421   &   0.5   &   3   &   29.1   &   1.94   &   0.00   &   -- \\ 
\end{tabular}
\end{center}
\caption{List of 2D materials with a ferromagnetic ground state (within the PBE approximation) containing two elements. ID denotes the unique ICSD/COD identifier (materials from ICSD have ID $<10^6$) for the bulk parent material and $J$ is the nearest neighbor exchange interaction obtained from the energy mapping. $E_\mathrm{Gap}$ denotes the electronic (Kohn-Sham) band gap. $\Delta$ is the spin wave gap obtained from the anisotropy constants and positive values indicate an out-of-plane easy axis.}
\label{tab:fm_2el}
\end{table*}

\begin{table*}[tb]
  \begin{center}
\begin{tabular}{lr@{\hskip 0.2in}r@{\hskip 0.2in}r@{\hskip 0.2in}rrr@{\hskip 0.2in}rr}
  {\bf Stoichiometry} & {\bf ID} & $S$ [$\hbar$] & $N_{nn}$ & $J$ [meV] & $\Delta$ [meV] & $E_\mathrm{Gap}$ [eV] & $T_\mathrm{C}$ \\
\hline
MnSiCa   &   1539705   &   1.0   &   4   &   3.63   &   0.587   &   0.00   &   29 \\ 
MnGeMg   &   1539696   &   1.0   &   4   &   11.1   &   0.956   &   0.00   &   75 \\ 
VClO   &   2106692   &   1.0   &   6   &   5.04   &   -0.0668   &   0.00   &   0 \\ 
VBrO   &   1537583   &   1.0   &   6   &   6.64   &   0.478   &   0.00   &   62 \\ 
CrClO   &   28318   &   1.5   &   6   &   1.04   &   0.0533   &   0.65   &   17 \\ 
CrBrO   &   1534386   &   1.5   &   6   &   0.337   &   -0.0607   &   0.50   &   0 \\ 
CrBrS   &   69659   &   1.5   &   6   &   6.03   &   -0.0884   &   0.46   &   0 \\ 
FeFO   &   291415   &   0.5   &   6   &   6.24   &   -0.616   &   0.00   &   0 \\ 
FeClO   &   2106381   &   0.5   &   6   &   7.74   &   0.377   &   0.00   &   -- \\ 
YClO$_2$   &   20449   &   0.5   &   2   &   126   &   -0.0153   &   0.00   &   0 \\ 
VOBr$_2$   &   24381   &   0.5   &   4   &   22.7   &   0.336   &   0.00   &   -- \\ 
TiKS$_2$   &   641335   &   0.5   &   6   &   10.4   &   0.00248   &   0.00   &   -- \\ 
TiRbS$_2$   &   77990   &   0.5   &   6   &   18.4   &   -0.00504   &   0.00   &   0 \\ 
CrHO$_2$   &   9012135   &   1.5   &   6   &   2.37   &   0.227   &   0.46   &   45 \\ 
CrPSe$_3$   &   626521   &   1.5   &   3   &   10.2   &   -0.0739   &   0.45   &   0 \\ 
CrSiTe$_3$   &   626810   &   1.5   &   3   &   3.36   &   0.0170   &   0.53   &   15 \\ 
CrGeTe$_3$   &   1543733   &   1.5   &   3   &   5.95   &   -0.370   &   0.36   &   0 \\ 
FeTaTe$_3$   &   2002027   &   1.0   &   2   &   48.5   &   1.49   &   0.00   &   0 \\ 
MnSeO$_4$   &   1527676   &   1.5   &   4   &   11.4   &   -0.401   &   0.02   &   0 \\ 
MoPO$_5$   &   36095   &   0.5   &   4   &   0.577   &   0.187   &   1.04   &   -- \\ 
CrMoF$_6$   &   50507   &   1.0   &   4   &   7.84   &   23   &   0.00   &   126 \\ 
CoCa$_2$O$_3$   &   1531759   &   1.5   &   4   &   2.94   &   2.57   &   0.03   &   67 \\ 
CrGa$_2$S$_4$   &   626052   &   2.0   &   6   &   1.88   &   0.0395   &   0.00   &   39 \\ 
CrGa$_2$Se$_4$   &   2001932   &   2.0   &   6   &   5.38   &   0.217   &   0.00   &   132 \\ 
NiV$_2$Se$_4$   &   1008112   &   1.0   &   6   &   24.0   &   0.153   &   0.00   &   122 \\ 
CrTa$_2$O$_6$   &   1001053   &   1.0   &   4   &   8.08   &   -0.693   &   0.24   &   0 \\ 
CuI$_2$O$_6$   &   4327   &   0.5   &   2   &   7.16   &   -0.0812   &   0.75   &   0 \\ 
CuV$_2$O$_6$   &   21067   &   0.5   &   2   &   1.97   &   -0.968   &   0.19   &   0 \\ 
SrTa$_2$O$_7$   &   154177   &   0.5   &   8   &   17.2   &   -3.17   &   0.00   &   0 \\ 
NiRe$_2$O$_8$   &   51016   &   1.0   &   6   &   1.50   &   2.56   &   1.58   &   31 \\ 
MnRe$_2$O$_8$   &   51014   &   0.5   &   6   &   1.35   &   -11.1   &   0.00   &   0 \\ 
CoCl$_2$O$_8$   &   33288   &   1.5   &   6   &   0.244   &   0.622   &   0.72   &   10 \\ 
NiCl$_2$O$_8$   &   33289   &   1.0   &   6   &   0.472   &   -0.155   &   1.54   &   0 \\ 
Fe$_2$In$_2$Se$_5$   &   155025   &   0.5   &   9   &   0.867   &   0.711   &   0.00   &   -- \\ 
Mn$_3$Cd$_2$O$_8$   &   1528776   &   0.5   &   4   &   0.0625   &   0.470   &   1.23   &   -- \\ 

\hline

CdGaInS$_4$   &   1538374   &   0.5   &   6   &   13.2   &   -0.0900   &   0.00   &   0 \\ 
VAgP$_2$Se$_6$   &   1509506   &   1.0   &   6   &   2.14   &   0.200   &   0.34   &   21 \\ 
CrCuP$_2$S$_6$   &   1000355   &   1.5   &   6   &   1.15   &   -0.0809   &   1.08   &   0 \\ 
NiLiP$_2$S$_6$   &   1541091   &   0.5   &   6   &   7.90   &   0.0930   &   0.00   &   -- \\ 
CoS$_2$C$_2$N$_2$   &   4330304   &   0.5   &   2   &   25.7   &   0.0765   &   0.00   &   -- \\ 
NiS$_2$C$_2$N$_2$   &   31320   &   1.0   &   2   &   6.86   &   -0.0635   &   0.65   &   0 \\ 
Mn$_2$FeC$_6$N$_6$   &   417824   &   2.5   &   3   &   0.102   &   0.155   &   1.83   &   4.4 \\ 

\hline

MnNa$_2$F$_3$P$_2$O$_7$   &   7022080   &   1.0   &   2   &   11.0   &   0.182   &   0.20   &   0 \\ 
VFC$_4$H$_4$O$_6$   &   2014296   &   1.0   &   2   &   22.3   &   0.0215   &   0.00   &   0 \\ 
CoC$_4$H$_8$N$_2$O$_4$   &   4509074   &   0.5   &   2   &   5.43   &   -0.379   &   0.59   &   0 \\ 
NiCl$_2$C$_6$H$_4$N$_2$   &   7227895   &   1.0   &   2   &   9.14   &   -0.505   &   0.85   &   0 \\ 
CuC$_6$H$_4$N$_6$O$_2$   &   7018416   &   0.5   &   2   &   3.04   &   0.0288   &   0.80   &   -- \\ 

\end{tabular}
\end{center}
\caption{List of 2D materials with a ferromagnetic ground state (within the PBE approximation) containing more than two elements. ID denotes the unique ICSD/COD identifier (materials from ICSD have ID $<10^6$) for the bulk parent material and $J$ is the nearest neighbor exchange interaction obtained from the energy mapping. $E_\mathrm{Gap}$ denotes the electronic (Kohn-Sham) band gap. $\Delta$ is the spin wave gap obtained from the anisotropy constants and positive values indicate an out-of-plane easy axis.}
\label{tab:fm_more_than_2}
\end{table*}

\newpage
\newpage
\begin{table*}
  \begin{center}
\begin{tabular}{lr@{\hskip 0.2in}r@{\hskip 0.2in}r@{\hskip 0.2in}rrr@{\hskip 0.2in}rr}
  {\bf Stoichiometry} & {\bf ID} & $S$ [$\hbar$] & $N_{nn}$ & $J$ [meV] & $\Delta$ [meV] & $E_\mathrm{Gap}$ [eV] & $T_\mathrm{C}$ \\
    \hline
CrSe   &   162899   &   0.5   &   4   &   -107   &   16.9   &   0.00   &   -- \\ 
MnSe   &   162900   &   0.5   &   4   &   -59.2   &   -10.5   &   0.00   &   0 \\ 
FeSe   &   633480   &   0.5   &   4   &   -95.1   &   -17.2   &   0.00   &   0 \\ 
TiBr$_2$   &   1535971   &   1.0   &   6   &   -6.99   &   -3.07   &   0.00   &   0 \\ 
VTe$_2$   &   603582   &   0.5   &   6   &   -2.62   &   -2.54   &   0.00   &   0 \\ 
VCl$_2$   &   1528165   &   1.5   &   6   &   -4.85   &   1.29   &   1.36   &   119 \\ 
VBr$_2$   &   246906   &   1.5   &   6   &   -6.89   &   1.42   &   1.29   &   158 \\ 
CrSe$_2$   &   626718   &   1.0   &   6   &   -19.8   &   -5.33   &   0.00   &   0 \\ 
MnCl$_2$   &   9009130   &   2.5   &   6   &   -0.639   &   0.111   &   2.03   &   33 \\ 
MnBr$_2$   &   9009109   &   2.5   &   6   &   -0.576   &   0.322   &   1.84   &   40 \\ 
MnI$_2$   &   9009110   &   2.5   &   6   &   -0.590   &   -0.502   &   1.43   &   0 \\ 
FeO$_2$   &   9009104   &   1.0   &   6   &   -0.132   &   -0.297   &   0.00   &   0 \\ 
FeO$_2$*   &   9009154   &   1.0   &   6   &   -2.59   &   -2.93   &   0.00   &   0 \\ 
CoI$_2$   &   9009100   &   0.5   &   6   &   -20.4   &   14.7   &   0.00   &   -- \\ 
RuCl$_3$   &   20717   &   0.5   &   3   &   -0.0368   &   -3.01   &   0.00   &   0 \\ 
VF$_4$   &   1539645   &   0.5   &   4   &   -14.0   &   -2.42   &   0.79   &   0 \\ 
NbF$_4$   &   2241796   &   0.5   &   4   &   -90.0   &   7.94   &   0.26   &   -- \\ 
RuF$_4$   &   165398   &   1.0   &   4   &   -1.54   &   -3.87   &   0.75   &   0 \\ 
               
\hline

MnGeCa   &   1539711   &   1.5   &   4   &   -25.2   &   7.22   &   0.00   &   433 \\ 
MnGeSr   &   1539720   &   1.5   &   4   &   -29.9   &   7.12   &   0.00   &   491 \\ 
MnGeBa   &   1539729   &   2.0   &   4   &   -11.5   &   5.28   &   0.00   &   358 \\ 
MnSnCa   &   1539717   &   2.0   &   4   &   -15.6   &   8.14   &   0.00   &   501 \\ 
VOCl$_2$   &   24380   &   0.5   &   4   &   -38.3   &   -4.28   &   0.00   &   0 \\ 
CuSiO$_3$   &   89669   &   0.5   &   2   &   -8.66   &   0.644   &   0.59   &   -- \\ 
VPS$_3$   &   648076   &   1.0   &   3   &   -108   &   0.58   &   1.08   &   261 \\ 
MnPS$_3$   &   61391   &   2.0   &   3   &   -3.32   &   -3.36   &   0.27   &   0 \\ 
MnPSe$_3$   &   643239   &   2.5   &   3   &   -4.42   &   -5.95   &   0.96   &   0 \\ 
NiPS$_3$   &   657314     &   1.0   &   3   &   -32.4   &   -4.68   &   0.88   &   0 \\ 
NiPSe$_3$   &   646145   &   1.0   &   3   &   -31.8   &   -2.06   &   0.62   &   0 \\ 
VMoO$_5$   &   1535988   &   0.5   &   4   &   -63.2   &   4.45   &   0.92   &   -- \\ 
AgSnF$_6$   &   1509332   &   0.5   &   2   &   -1.16   &   1.71   &   0.61   &   -- \\ 
CrNbF$_6$   &   4030623   &   2.5   &   4   &   -4.15   &   -9.61   &   0.26   &   0 \\ 
CuLi$_2$O$_2$   &   174134   &   0.5   &   2   &   -4.81   &   -1.25   &   0.44   &   0 \\ 
MnGa$_2$S$_4$   &   634670   &   2.0   &   6   &   -13.4   &   4.22   &   0.00   &   549 \\ 
MnAl$_2$S$_4$   &   608511   &   2.0   &   6   &   -18.0   &   4.67   &   0.00   &   702 \\ 
MnIn$_2$Se$_4$   &   639980   &   2.5   &   6   &   -3.03   &   1.70   &   0.00   &   209 \\ 
NiGa$_2$S$_4$   &   634901   &   1.0   &   6   &   -12.3   &   -2.76   &   0.15   &   0 \\ 
MnBi$_2$Te$_4$   &   7210230   &   2.5   &   6   &   -0.35   &   0.852   &   0.71   &   35 \\ 
ReAg$_2$Cl$_6$   &   4508861   &   1.5   &   6   &   -3.42   &   7.93   &   1.01   &   143 \\ 
CoRe$_2$O$_8$   &   51015   &   1.5   &   6   &   -2.22   &   6.39   &   0.59   &   98 \\ 
MnSb$_2$F$_12$   &   1535152   &   2.5   &   6   &   -0.292   &   0.251   &   1.95   &   22 \\ 
Mn$_2$Ga$_2$S$_5$   &   634664   &   2.5   &   9   &   -4.15   &   0   &   0.23   &   0 \\ 
Fe$_2$Ga$_2$S$_5$   &   631804   &   1.0   &   9   &   -0.00284   &   -0.364   &   0.00   &   0 \\ 

\hline

MnSbClS$_2$   &   151925   &   2.5   &   2   &   -4.02   &   -2.70   &   0.34   &   0 \\ 
MnSbBrS$_2$   &   1528449   &   2.5   &   2   &   -3.83   &   -2.66   &   0.10   &   0 \\ 
MnSbBrSe$_2$   &   1528451   &   2.5   &   2   &   -4.23   &   -3.18   &   0.30   &   0 \\ 
MnSbISe$_2$   &   2013470   &   2.5   &   2   &   -4.15   &   -4.83   &   0.00   &   0 \\ 
FeMoClO$_4$   &   1530888   &   2.5   &   4   &   -0.461   &   -3.13   &   1.36   &   0 \\ 
FeWClO$_4$   &   80798   &   2.5   &   4   &   -0.421   &   -3.33   &   1.56   &   0 \\ 
MnMoTeO$_6$   &   291413   &   2.5   &   4   &   -1.51   &   -1.32   &   1.59   &   0 \\ 
FeBrSr$_2$O$_3$   &   7221295   &   2.0   &   4   &   -0.153   &   0.614   &   0.00   &   8 \\ 
MnSH$_2$O$_4$   &   74810   &   2.5   &   4   &   -0.725   &   0.187   &   2.52   &   28 \\ 
CoSeH$_2$O$_4$   &   408100   &   1.5   &   4   &   -2.00   &   1.15   &   0.65   &   41 \\ 
VAgP$_2$S$_6$   &   1509505   &   1.0   &   2   &   -1.75   &   -1.38   &   0.17   &   0 \\ 
CuPtC$_3$N$_4$   &   1534876   &   1.5   &   9   &   -0.0457   &   0   &   1.68   &   0 \\ 
MnSe$_2$C$_6$N$_4$   &   7112837   &   1.0   &   2   &   -33.1   &   12.4   &   0.00   &   0 \\ 
Fe$_2$Br$_2$Sr$_3$O$_5$   &   1529142   &   2.0   &   5   &   -29.4   &   -8.62   &   0.00   &   0 \\ 

\hline

OsF$_5$KMO   &   166586   &   0.5   &   2   &   -0.421   &   2.88   &   0.91   &   -- \\ 
CoPO$_4$CH$_3$   &   1528341   &   1.5   &   4   &   -2.03   &   3.94   &   0.74   &   56 \\ 
CoCl$_2$C$_4$H$_4$N$_2$   &   7218183   &   0.5   &   2   &   -23.2   &   -2.83   &   0.47   &   0 \\ 
NiC$_2$O$_4$C$_2$H$_8$N$_2$   &   4509073   &   1.0   &   2   &   -16.9   &   1.92   &   1.80   &   0 \\ 
\end{tabular}
\end{center}
\caption{List of 2D materials with an anti-ferromagnetic ground state (within the PBE approximation). ID denotes the unique ICSD/COD identifier (materials from ICSD have ID $<10^6$) for the bulk parent material and $J$ is the nearest neighbor exchange interaction obtained from the energy mapping. $E_\mathrm{Gap}$ denotes the electronic (Kohn-Sham) band gap. $\Delta$ is the spin wave gap obtained from the anisotropy constants and positive values indicate an out-of-plane easy axis.}
\label{tab:afm}
\end{table*}

\newpage

\begin{table}[tb]
  \begin{center}
\begin{tabular}{lrr}
 {\bf Formula}  &  ID & Comment \\
    \hline
TiBr$_3$   &   1535636 & AFM configuration unstable \\ 
Ta$_2$SrO$_7$   &   154177 & AFM configuration unstable \\ 
YClO$_2$   &   20449 & AFM configuration unstable \\ 
V$_2$LiO$_5$   &   88640 & No simple AFM configurations \\ 
V$_2$H$_2$O$_5$   &   260368 & No simple AFM configurations \\ 
Ta$_2$BaO$_7$   &   1526608 & No simple AFM configurations \\  
Ni$_2$As$_2$O$_7$   &   2104863 & Vertical dimer \\ 
CoSb$_2$Br$_2$O$_3$   &   418858 & Vertical dimer \\ 
Nb$_3$Cl$_8$   &   408645 & Trimer \\ 
Nb$_3$Br$_8$   &   1539108 & Trimer \\ 
Nb$_3$I$_8$   &   1539109 & Trimer \\ 
\end{tabular}
\end{center}
\caption{List of 2D ferromagnetic compounds, which did not allow for a simple estimation of a nearest neighbor exchange coupling constant.}
\label{tab:failed}
\end{table}

In Tab. \ref{tab:fm_2el} we list all the predicted ferromagnetic insulators containing two elements and tn Tab. \ref{tab:fm_more_than_2} we list the ferromagnetic materials containing three, four or five elements. For all materials we provide the COD/ICSD identifier for the bulk parent compound from which the 2D material was derived. We also stats the spin $S$, the number of nearest neighbors $N_{nn}$, the exchange coupling $J$, the spinwave gap $\Delta$, and Kohn-Sham band gap $E_\mathrm{Gap}$. For materials with $S\neq1/2$ and $N_{nn}\neq2$ we have calculated an estimated critical temperature from Eq. \eqref{eq:tc}. In Tab. \ref{tab:afm} we show all the anti-ferromagnetic compounds found in the computational screening. In addition, we found 11 materials for which we were not able to evaluate exchange coupling constants. This was either due to problems converging the anti-ferromagnetic spin configuration (converged to ferromagnetic state), more than two magnetic atoms in the unit cell, or that the two magnetic atoms in the unit cell form a vertical dimer. All of the materials are, however, predicted to be magnetic and could comprise interesting magnetic 2D materials that are exfoliable from 3D parent compounds.

\bibliography{references.bib}

\end{document}